\documentclass[12pt,preprint]{aastex}
\usepackage{emulateapj5}
\usepackage{apjfonts}
\usepackage{epsfig}
\usepackage{natbib}
\newcommand{\aox}{\ensuremath{\alpha_{\mathrm{ox}}}}
\newcommand{\asca}{\emph{ASCA}}

\newcommand{\civ}{\ion{C}{4}}
\newcommand{\chisq}{\ensuremath{\chi^2}}
\newcommand{\chisqr}{\ensuremath{\chi^2_r}}
\newcommand{\chandra}{\emph{Chandra}}

\def\gtrsim{\mathrel{\hbox{\rlap{\hbox{\lower4pt\hbox{$\sim$}}}\hbox{\raise2pt\hbox{$>$}}}}}

\newcommand{\fwhb}{\ensuremath{\mathrm{FWHM}_\mathrm{H{\beta}}}}

\newcommand{\halpha}{{\rm H}\ensuremath{\alpha}}

\newcommand{\hbeta}{{\rm H}\ensuremath{\beta}}

\newcommand{\hst}{\emph{HST}}

\newcommand{\kms}{km~s\ensuremath{^{-1}}}

\newcommand{\lf}{\ensuremath{L_{\rm{5100 \AA}}}}

\newcommand{\lledd}{\ensuremath{L_{\mathrm{bol}}/L{\mathrm{_{Edd}}}}}

\newcommand{\loiii}{\ensuremath{L_{\mathrm{[O~ {\tiny III}]}}}}
\newcommand{\loiv}{\ensuremath{L_{\mathrm{[O~ {\tiny IV}]}}}}

\newcommand{\lhb}{\ensuremath{L_{\mathrm{H{\beta}}}}}

\newcommand{\lx}{\ensuremath{L_{\mathrm{X}}}}
\newcommand{\lhard}{\ensuremath{L_{\mathrm{2-10~ keV}}}}

\newcommand{\mbh}{\ensuremath{M_\mathrm{BH}}}

\newcommand{\msigma}{\ensuremath{M_{\mathrm{BH}}-\sigmastar}}
\newcommand{\msun}{\ensuremath{{\rm M}_{\odot}}}

\newcommand{\oiii}{[\ion{O}{3}]}
\newcommand{\oiv}{[\ion{O}{4}]}

\newcommand{\rblr}{\ensuremath{R_{\mathrm{BLR}}}}

\newcommand{\sigmastar}{\ensuremath{\sigma_{\ast}}}

\newcommand{\swift}{\emph{Swift}}

\newcommand{\xmm}{{\it XMM-Newton}}

\def\lax{{$\mathrel{\hbox{\rlap{\hbox{\lower4pt\hbox{$\sim$}}}\hbox{$<$}}}$}}
\def\gax{{$\mathrel{\hbox{\rlap{\hbox{\lower4pt\hbox{$\sim$}}}\hbox{$>$}}}$}}

\slugcomment{Spet 2, 2010;
to be published by {\it The Astrophysical Journal}.}
\shorttitle{{\it RL}}
\shortauthors{GREENE ET AL.}

\begin{document}

\title{The Lick AGN Monitoring Project: Alternate Routes to a Broad-line 
Region Radius}

\author{Jenny E. Greene\altaffilmark{1}, Carol E. Hood\altaffilmark{2}, 
Aaron J. Barth\altaffilmark{2}, Vardha N. Bennert\altaffilmark{3},
Misty C. Bentz\altaffilmark{2,4}, Alexei V. Filippenko\altaffilmark{5},
Elinor Gates\altaffilmark{6},
Matthew A. Malkan\altaffilmark{7},
Tommaso Treu\altaffilmark{3,8}, Jonelle L. Walsh\altaffilmark{2}, and
Jong-Hak Woo\altaffilmark{9}}

\altaffiltext{1}{Department of Astrophysical Sciences, Princeton University, 
Princeton, NJ 08544; Princeton-Carnegie Fellow.}

\altaffiltext{2}{Department of Physics and Astronomy, 
4129 Frederick Reines Hall, University of California, Irvine, CA 92697-4575.}

\altaffiltext{3}{Department of Physics, University of California, Santa 
Barbara, CA 93106.}

\altaffiltext{4}{Hubble Fellow.}

\altaffiltext{5}{Department of Astronomy, University of California, Berkeley, 
CA 94720-3411.}

\altaffiltext{6}{Lick Observatory, P.O. Box 85, Mount Hamilton, CA 95140.}

\altaffiltext{7}{Department of Physics and Astronomy, University of 
California, Los Angeles, CA 90024.}

\altaffiltext{8}{Sloan Fellow; Packard Fellow.}

\altaffiltext{9}{Astronomy Program, Department of Physics and Astronomy, 
Seoul National University, Seoul, Korea, 151-742.}

\begin{abstract}

It is now possible to estimate black hole masses across cosmic time,
using broad emission lines in active galaxies.  This technique informs our
views of how galaxies and their central black holes coevolve.
Unfortunately, there are many outstanding uncertainties associated
with these ``virial'' mass estimates.  One of these comes from using
the accretion luminosity to infer a size for the broad-line
region.  Incorporating the new sample of low-luminosity active
galaxies from our recent monitoring campaign at Lick Observatory, we
recalibrate the radius-luminosity relation with tracers of the
accretion luminosity other than the optical continuum.  We find that the 
radius of the broad-line region scales as the square root of the X-ray
and \hbeta\ luminosities, in agreement with recent optical studies.
On the other hand, the scaling appears to be marginally steeper with
narrow-line luminosities.  This is consistent with a previously
observed decrease in the ratio of narrow-line to X-ray luminosity
with increasing total luminosity.  The radius of the broad-line region 
correlates most tightly with \hbeta\ luminosity, while the X-ray and
narrow-line relations both have comparable scatter of a factor of
two.  These correlations provide useful alternative virial BH masses
in objects with no detectable optical/UV continuum emission, such as
high-redshift galaxies with broad emission lines, radio-loud
objects, or local active galaxies with galaxy-dominated continua.

\end{abstract}

\keywords{galaxies: active --- galaxies: nuclei --- galaxies: Seyfert} 

\section{The Radius-Luminosity Relation}

Over the past decade, interest in measuring supermassive black hole
(BH) masses has intensified, as evidence mounts that BHs play a
central role in galaxy evolution
\citep[e.g.,][]{silkrees1998,hopkinsetal2006}.  Locally, BH masses are
measured using stars, gas disks, or megamaser disks as dynamical
tracers \citep[e.g.,][]{gultekinetal2009}.  None of these techniques
can currently reach beyond a few tens of Mpc.  Thus, we
resort to indirect mass estimates in actively accreting BHs to probe
BH and galaxy coevolution at cosmological distances.  Studies of the
BH mass and accretion-rate distributions both locally
\citep{greeneho2007b,schulzewisotzki2010} and at higher redshifts
\citep{wooetal2006,kollmeieretal2006,shenetal2008b,wooetal2008,
vestergaardosmer2009,kellyetal2009},
as well as studies of possible evolution in BH-bulge scaling relations
\citep[e.g.,][]{shieldsetal2003,treuetal2004,walteretal2004,pengetal2006a,
  pengetal2006b,treuetal2007,salvianderetal2007,alexanderetal2008,
  jahnkeetal2009,greeneetal2010,bennertetal2010}, all rely on BH masses
derived from active galactic nuclei (AGNs).

BH masses derived from AGNs use the broad-line region (BLR) gas as the
dynamical tracer, based on the assumption that the gas is primarily
accelerated by the gravity of the BH.  The gas velocity dispersion is
derived from the broad-line width, but the BH mass estimate also
requires the radius of the emitting region. The best estimate for its
size comes from ``reverberation'' or echo mapping
\citep{blandfordmckee1982}.  Detailed spectroscopic monitoring allows
an estimate of the light-travel time through the BLR, by measuring the
delay between variations in the continuum and line emission \citep[see
the recent compilation by ][]{petersonetal2004}.  This technique has a
long history
\citep[e.g.,][]{antonuccicohen1983,petersonetal1983,ulrichetal1984,
  gaskellsparke1986}, and thus far has yielded reliable sizes for a
few dozen sources \citep[see][]{petersonetal2004}.

Five reverberation-mapped sources show a $1/\sqrt{R}$ decline in
velocity width ranging from \civ\ $\lambda$1549 to \hbeta, as expected
for a virialized BLR in a $1/R$ potential
\citep{kollatschny2003,petersonetal2004}.  Data from our Lick AGN
Monitoring Project (LAMP), the subject of this paper, are consistent
with the same assumption; when multiple Balmer lines are considered
independently (e.g., \halpha, \hbeta, H$\gamma$), all yield consistent
estimates for the so-called virial product, $\upsilon^2 R/G$
\citep{bentzetal2010lamp}.  On the other hand, other models, such as
disk winds, would predict similar radial dependence
\citep[e.g.,][]{murraychiang1995}.  The importance of radiation
pressure in supporting the BLR is currently a matter of debate as well
\citep[e.g.,][]{marconietal2008,marconietal2009,netzer2009,netzermarziani2010}.
Despite these major uncertainties, the reverberation-derived BH masses
correlate remarkably well with the luminosities and stellar velocity
dispersions of their host bulges
\citep{bentzetal2009mbul,wooetal2010}.  In addition, the very few existing
direct dynamical measures of BH masses have so far turned out to be
consistent with the reverberation-mapping virial estimates
\citep{daviesetal2006,onkenetal2007,hicksmalkan2008}.

Since reverberation radii are usually not available, a secondary
estimate of BLR size is often obtained from the empirical
correlation (the ``radius-luminosity'' relation) between AGN
luminosity and BLR size, $R_{\rm BLR} \propto L^{\beta}$
\citep{kaspietal2000,kaspietal2005,bentzetal2006,bentzetal2009rl}.
With just a measurement of the AGN luminosity, typically \lf, and a
broad-line width, typically \fwhb, one can roughly estimate a BH mass
as \mbh$=f\,\upsilon^2 L^{\beta}/G$. Here $f$ is a scaling parameter
that includes unknown information about the geometry and kinematics of
the BLR.

These so-called ``single-epoch'' virial BH masses are indirect, and
depend on a number of assumptions.  Two empirically determined
quantities fundamentally limit the accuracy of the derived BH masses.
One is $f$, which is currently determined
for ensembles of active galaxies through comparison between AGN-based
masses and other estimates of \mbh\ such as the \msigma\ relation
\citep[e.g.,][]{gebhardtetal2000b,
  ferrareseetal2001,nelsonetal2004,onkenetal2004,
  greeneho2006msig,shenetal2008a, wooetal2010}.  While there are good
reasons to suspect that $f$ may depend on physical properties of the
BH such as accretion rate \citep[e.g.,][]{collinetal2006},
reverberation-mapping campaigns have not yet succeeded in measuring
$f$ directly for individual objects.  We are getting closer, however, 
since two-dimensional
reverberation mapping is growing more common and the velocity-resolved 
emission-line response strongly constrains $f$ in individual sources
\citep[e.g.,][]{kollatschny2003,bentzetal2008lamp,denneyetal20092d}.

The other empirically determined parameter is the slope of the
radius-luminosity relation, $\beta$, which is the subject of this
paper.  We are motivated to revisit this question thanks to our
recent reverberation-mapping campaign, which has doubled the number of
reverberation-mapped AGNs with $R_{\rm BLR} \lesssim 10$ light days.
We do not consider the optical AGN continuum luminosity because the
\emph{Hubble Space Telescope} (\hst) is required to spatially
disentangle the AGN and galaxy continuum for these low-luminosity
sources.  The requisite \hst\ imaging is underway (GO-11662, PI Bentz), 
and we will present
the optical radius-luminosity relation in a future paper.  Here
we focus on other direct and indirect indicators of the AGN
luminosity, including the X-ray luminosity and broad and narrow
emission-line luminosities.

There are practical reasons to consider other routes to determining
BLR radii.  For example, alternate relations are useful whenever the
optical/UV continuum from the AGN is not measurable.  This could occur
when the AGNs are radio loud, so that the optical/UV continuum is
contaminated by synchrotron radiation, or when the galaxy rather than
the AGN dominates the optical continuum
\citep[e.g.,][]{wuetal2004,greeneho2005cal}. It has become common
practice to use \halpha\ or \hbeta\ luminosities to calculate \rblr\
for high-redshift targets where the continuum is rarely detected
\citep[e.g.,][]{alexanderetal2008,shapiroetal2009}.  Finally,
remarkably, there is indirect evidence that broad-line widths measured
from polarized line emission may provide a reasonable single-epoch
virial BH mass \citep[e.g.,][C.~Y. Kuo, in
preparation]{zhangetal2008pol,liuetal2009}.  In these cases, hard
X-rays or narrow emission lines are some of the only available proxies
for AGN continuum luminosity.

\psfig{file=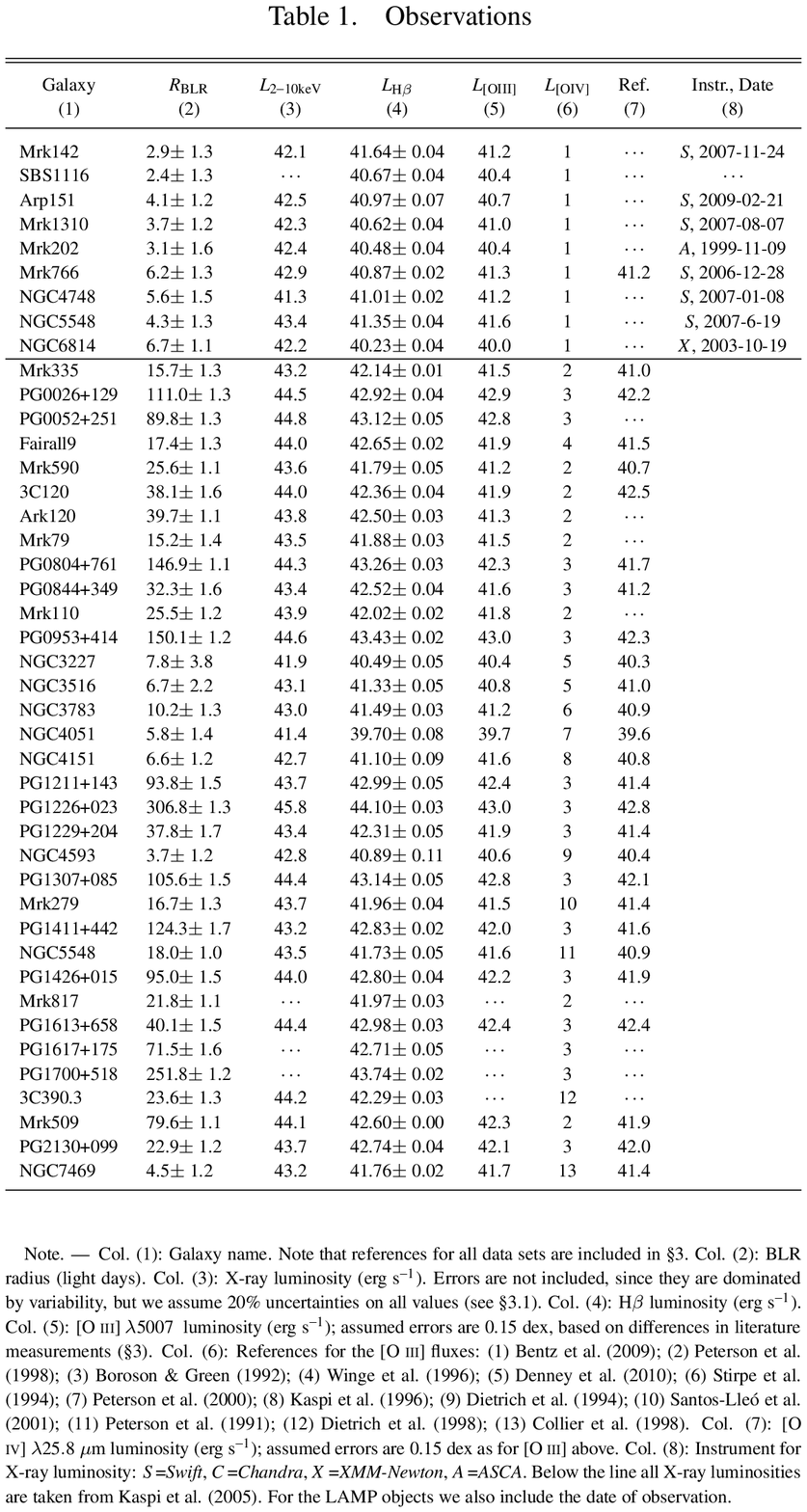,width=0.4\textwidth,keepaspectratio=true,angle=0}
\vskip 4mm

\subsection{Which Luminosity Best Predicts BLR Size?}

If the spectral energy distributions (SEDs) and the density
distribution in the BLR are independent of luminosity, then we expect
that the BLR size will scale simply with the square-root of
the photoionizing luminosity, $R_{\rm BLR} \propto \sqrt{L}$
\citep[e.g.,][]{netzer1990}.   The most recent calibrations of the
radius-luminosity relation have all been consistent with this simple
relation \citep{bentzetal2006,bentzetal2009rl}.  

If the SEDs were really independent of luminosity, the photoionizing
luminosity could be estimated from a measure of the AGN continuum at
almost any wavelength.  However, there are observational indications
of luminosity dependence in SEDs.  Indeed, on theoretical grounds we
might also expect smaller and hotter accretion disks around lower-mass 
BHs \citep[e.g.,][]{shields1978,zhengmalkan1993}.  While the equivalent
width of \hbeta\ is constant in high-luminosity active galaxies
\citep[e.g.,][]{searlesargent1968}, both \citet{croometal2002} and
\citet{greeneho2005cal} see evidence for a weak inverse-Baldwin effect
in \hbeta\ at low luminosity.  Furthermore, the increase in \aox\ with
UV luminosity \citep[e.g.,][]{avnitananbaum1982,steffenetal2006,
  desrochesetal2009} suggests luminosity-dependent changes in the SED.
Finally, the relative strengths of the ``big blue bump'' and the
X-rays depend on Eddington ratio, with the latter dominating at lower
\lledd\ \citep[][]
{malkansargent1982,vasudevanfabian2007,vasudevanfabian2009}.  For a
review of the situation at yet lower \lledd\ see \citealt{ho2008}.
Therefore, in this paper we will consider several observables which
may correlate with, and thus be used to estimate, the ionizing
luminosity.  Specifically, we consider the following proxies for the
AGN luminosity: hard X-ray luminosity ($L_{\rm 2-10~ keV}$), \hbeta\
luminosity (\lhb), narrow \oiii$~\lambda 5007$~\AA\ luminosity
(\loiii), and narrow \oiv\ $\lambda$25.8 $\mu$m luminosity (\loiv).

Throughout we assume the following cosmological parameters to calculate
distances: $H_0 = 100\,h = 70$~\kms~Mpc$^{-1}$, $\Omega_{\rm m} = 0.30$,
and $\Omega_{\Lambda} = 0.70$.

\section{The Lick AGN Monitoring Project}

The new measurements that motivate this work result from LAMP, a
dedicated monitoring campaign of 13 AGNs (including the well-studied
Seyfert galaxy NGC 5548). We specifically focused on nearby (redshift
$z < 0.05$) Seyfert galaxies with low luminosities ($\lambda L_{\rm
  5100 \AA} \lesssim 10^{43}$~erg~s$^{-1}$) and probable BH masses in
the range $10^6 - 3 \times 10^7$~\msun, since this luminosity and mass
regime had not been explored fully in the past.  Spectroscopic
monitoring was carried out with the Lick Observatory 3 m Shane
telescope over a nearly contiguous 64 day period
\citep{bentzetal2009lamp}, while photometric monitoring was performed
over a longer period utilizing four smaller telescopes
\citep{walshetal2009}.  We successfully measured BLR radii based on
\hbeta\ for nine objects \citep{bentzetal2009lamp}, reported lags in
multiple other Balmer transitions \citep{bentzetal2010lamp}, and
succeeded in measuring velocity-resolved lags in at least three
sources \citep{bentzetal2008lamp,bentzetal2009lamp}.  Finally, we
revisited the calibration of reverberation-mapped BH masses using the
\msigma\ relation \citep{wooetal2010}.  For the purpose of this paper,
we focus on BLR radii based exclusively on \hbeta\ lag times.

\section{Luminosities and BLR Radii}

The BLR light-crossing times used here are presented by
\citet{bentzetal2009rl} and \citet{bentzetal2009lamp} for the previous
reverberation-mapped and LAMP AGNs, respectively. We note that improved 
lag measurements were more recently reported for a subset of galaxies by
\citet{denneyetal2010}.  We have confirmed that the radius-luminosity
relation based on \hbeta\ does not change with the inclusion of their
lag values, but continue to use the old measurements for temporal
consistency with the X-ray observations.  We follow
\citet{bentzetal2009rl} and \citet{petersonetal2004} and remove IC
4329A from the sample due to uncertainties in the measurements.
Throughout we will refer to the sample of active galaxies with
reverberation mapping, excluding the LAMP targets, as the ``non-LAMP''
objects.  We describe the origin of the AGN luminosities in this
section (Table 1).  It is useful to remember that ``BLR size'' here
actually refers to the time of peak response of the \hbeta\ emitting 
gas relative to the continuum, multiplied by the speed of light.  Had the
experiment been done with C {\small IV} $\lambda$1549, for instance, 
the sizes would have been smaller, but the widths larger.  When 
calculating effective BH masses, it is important to match the species 
used to measure velocity dispersion with the radius relation 
calibrated for the same species.

Of all the luminosities we discuss, only the broad \hbeta\ line
luminosity (\lhb) is measured as part of the reverberation-mapping
campaign, simultaneously with the radius measurement.  It should
provide a fairly direct and unbiased probe of the photoionizing
continuum.  Both \citet{bentzetal2009lamp} and \citet{kaspietal2005}
tabulate average \lhb\ measured in the same way.  
\vbox{ 
\hskip 0.2in
\psfig{file=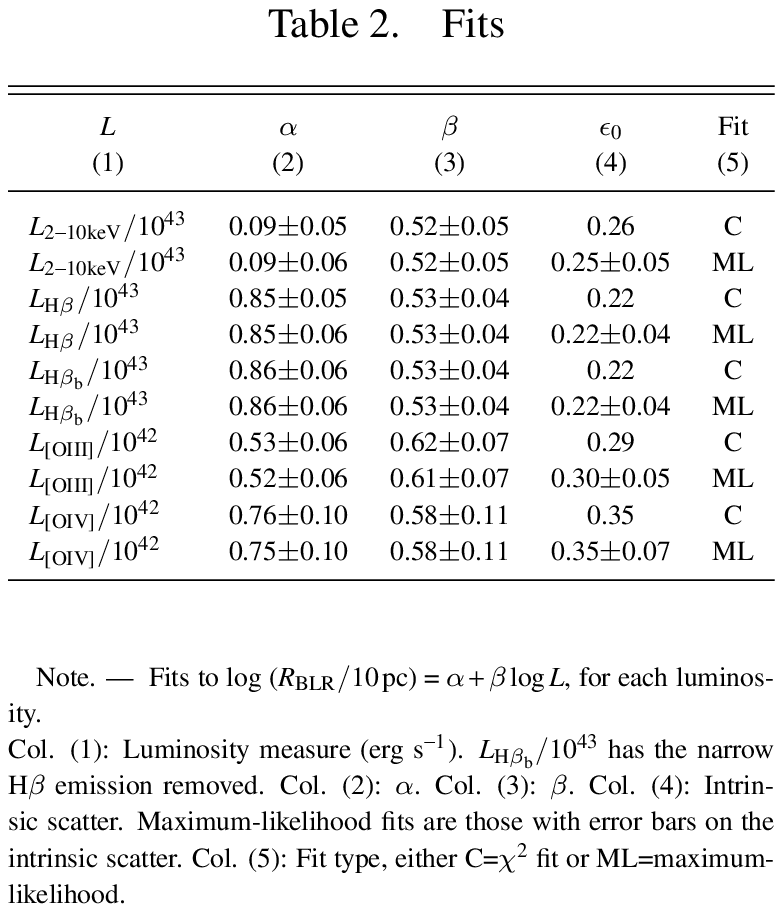,width=0.4\textwidth,keepaspectratio=true,angle=0}
}
\vskip 4mm
\noindent
Note that we present results based both on the total \hbeta\
luminosity (narrow and broad combined) and the broad \hbeta\
luminosity alone.  The results are basically identical, since the
median luminosity difference is less than 5\%. Although it would be
useful to examine \halpha\ as well, uniform measurements do not exist
for the non-LAMP sample, and thus we must await future work.

The \oiii\ luminosities for the LAMP sample itself are measured from
the Shane spectra and are presented by \citet{bentzetal2009lamp}.  For
the non-LAMP targets, we draw from previous reverberation mapping
campaigns for the local galaxies. For the more distant and
luminous Palomar-Green quasars \citep{schmidtgreen1983}, we combine
the equivalent-width measurements of \oiii\ from
\citet{borosongreen1992} with the continuum fluxes of 
\citet{kellermanetal1989} as given by \citet{hopeng2001} and
\citet{greeneetal2006}.  Table 1 contains all measurements, including
relevant references.  Note that the \oiii\ luminosities have not been
corrected for extinction.  The formal errors for the \oiii\ 
measurements are in the range $2-15\%$.  However, we find a median 
difference of $\sim 40\%$ between different literature values 
\citep[relying predominantly on the compilation of][]{whittle1992a}. 
The values used here, from previous reverberation mapping campaigns, 
are typically smaller than those compiled by Whittle.
Thus, in our fitting we adopt an uncertainty of 0.15 dex in the line 
luminosities, as an estimate of the impact of various systematic 
effects discussed below.

A large fraction of the non-LAMP sources have \oiv\ $\lambda$25.8 
$\mu$m luminosities
available in the literature from the {\it Spitzer Space Telescope}. We
draw predominantly from the measurements of \citet{dasyraetal2008},
which cover a large fraction of the non-LAMP reverberation-mapped
sample.  We take the measurements for Mrk 766 and Mrk 335 from the
work of \citet{tommasinetal2010} and that of NGC 3516 from
the work of \citet{gallimoreetal2010}.  With the exception of this
last, all were taken with the high-resolution grating.  A similar
comparative exercise as above, this time with the compilations of
\citet{tommasinetal2010}, \citet{gallimoreetal2010}, and
\citet{veilleuxetal2009}, yields a lower uncertainty estimate 
($\sim 10\%$) than for the \oiii\ lines. Presumably greater
agreement is reached because the data sets were in many cases
identical, and so we use a value of $40\%$ as above.  

The X-ray luminosities for the non-LAMP sources are taken directly
from the compilation of \citet{kaspietal2005}.  They are derived from
a variety of literature sources, but generally are based on {\it Advanced
Satellite for Cosmology and Astrophysics} (\asca) data
\citep{tanakaetal1994}.  While many of these targets have more recent
\xmm\ or \chandra\ observations available in the archive, the \asca\
measurements are actually closer in time to the reverberation-mapping
campaign.  Thus, we adopt the X-ray luminosities presented by 
Kaspi et al. in all cases.

The X-ray luminosities for the LAMP objects come from a range of
sources.  In all cases we adopt the observation closest in time to
that of our campaign (Spring 2008).  Six objects (Mrk 142, Apr 151,
Mrk 766, Mrk 1310, NGC 5548, and NGC 4748) have X-ray luminosities from
\swift.  All were observed between 2007 and 2009.  The \swift\ data
were extracted using the {\it xselect} task as part of the {\tt
  HEASARC}\footnotetext{http://heasarc.gsfc.nasa.gov/} 
tool-set.  Each source was extracted from a circular region
with a radius of 20 pixels ($\sim 47\arcsec$).  Background rates were
negligible in all cases. Count rates were converted to fluxes assuming
a power-law spectrum with $\Gamma=1.8$ ($E \propto E^{-\Gamma}$) and
no internal absorption.  In the few cases with multiple epochs (e.g., Mrk
766 and NGC 5548), we analyze the longest observation where the galaxy
center is close to the image center.

The remainder of the LAMP AGNs only have heterogeneous measurements
available in the literature. In one case (NGC 6814) we use the \xmm\
slew survey \citep{saxtonetal2008} and in another (Mrk 202) we resort
to an \asca\ observation from 1999 \citep{uedaetal2005}, where
aperture photometry yields a count rate that is converted to a flux
assuming only Galactic extinction and $\Gamma=1.7$.  The luminosity for
NGC 6814, from the slew survey, was derived in a similar fashion,
using the same spectral model.  The only difference is that the flux is
reported for 2--12 keV. We use
webPIMMS\footnote{http://heasarc.gsfc.nasa.gov/Tools/w3pimms.html.} to
calculate the 2--10 keV flux assuming our spectral model.

With many years of comprehensive monitoring, NGC 5548 is a special
case and warrants extra attention.  There are fourteen non-LAMP epochs
from \citet{petersonetal2002} and \citet{bentzetal2007}.  In Figure 1b 
we show all fifteen epochs in gray for reference.
Currently, NGC 5548 is in a very low luminosity state.  The LAMP
measurement differs by a factor of $\sim 4$ from the weighted average
of all other epochs \citep[e.g.,][]{bentzetal2009lamp}.
Unfortunately, we have only two epochs of X-ray data for this source,
and, given narrow-line region sizes of hundreds of pc, 
the \oiii\ luminosity is presumed constant over timescales of 
months.  For the purposes of fitting, we adopt the weighted
average lag, $18 \pm 0.6$ light-days, from \citet{bentzetal2009rl} as
the non-LAMP point.  The early X-ray data are from \asca\ and were
taken in 1993, when the lag was measured to be $13^{+1.6}_{-1.4}$
light-days \citep{petersonetal2002}. If we rather adopt the latter
value in our fitting of the X-ray radius-luminosity relation (\S 4),
it makes no difference to our results.  Since the BLR size of NGC 5548 
has been observed to change on timescales short compared to 
changes in the narrow-line flux, it is also interesting to note that there is a
scatter of $0.2\pm0.1$ dex in the logarithm of the ratio of lag to
\oiii\ luminosity across the fifteen epochs.

\subsection{Systematics: X-ray Variability, Aperture Effects, and Extinction}

Each of the luminosities we consider comes with its own complications.
In the case of the narrow emission lines, they have been photoionized
by the average continuum luminosity over the past $\sim$ 100 years,
during which time \rblr\ may vary significantly.  On the other hand,
the X-ray emission region is more compact than the optical emitting
region, and thus varies on shorter timescales than changes in \rblr\
occur.  With nonsimultaneous observations, we may introduce
significant scatter into the \rblr--\lhard\ relation.  In addition,
X-ray variability timescales depend systematically on \mbh\ and
luminosity \citep[e.g.,][]{oneilletal2005,mchardyetal2006,
  miniuttietal2009}. Thus, it is at least conceivable that some
systematic bias is introduced into the \rblr--\lhard\ relation.  We
investigate that possibility here.

We start by considering all multi-epoch data available for
reverberation-mapped sources from the Tartarus
database\footnote{http://tartarus.gsfc.nasa.gov/.}.  The benefit of
Tartarus is that the fluxes have been derived from the \asca\ X-ray
spectra in a uniform way.  Spectral fits to the hard X-rays (2--10 keV)
are performed, with the region around Fe~K$\alpha$ masked and
including possible internal absorption (which is small in this
spectral region).  Ten of the non-LAMP targets have multiple epochs of
observations spanning more than one year in the Tartarus database.
They include 3C 120, Fairall 9, Mrk 509, NGC 3227, NGC 3516, NGC 3783,
NGC 4051, NGC 4151, NGC 4593, NGC 5548, NGC 4269, and PG 1226+023.
The typical cadence is a few observations per year.

For each object we calculate a mean and root-mean square (rms) flux using 
the Tartarus database.  We find variability amplitudes of 5--80\% (one
standard deviation) over the 1--7 yr timescales probed by these
observations.  NGC 3516 is the target with the highest variability
amplitude (80\%). The majority of objects do not vary even by a factor
of two on these timescales.  The median amplitude of variability is
$\sim 20\%$.  Thus, the level of intrinsic variability in the X-ray
luminosity of most Seyferts is usually too small to impact the
\rblr--\lhard\ relation.  For fitting purposes we thus adopt $20\%$
uncertainties in all X-ray fluxes.

\begin{figure*}
\vbox{ 
\vskip -8mm
\hskip 0.4in
\psfig{file=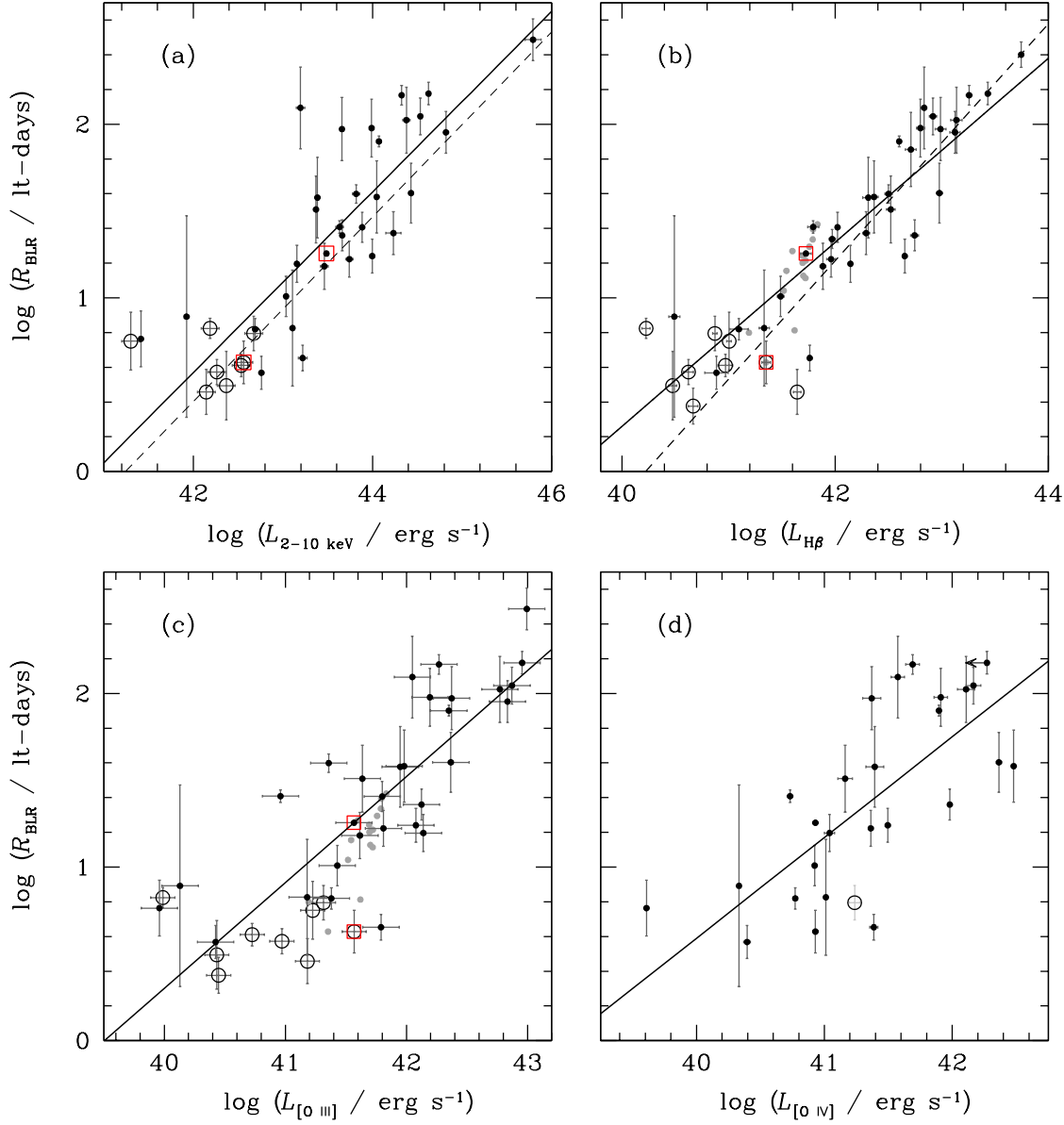,width=0.8\textwidth,keepaspectratio=true,angle=0}
}
\vskip -0mm

\figcaption[]{ 
({\bf a}) Fit to the $L_{\rm 2-10~ keV}-R_{\rm BLR}$ relation, including both 
LAMP sources ({\it open circles}) and non-LAMP sources ({\it filled dots}). 
Our maximum-likelihood fit ({\it solid}; $\beta=0.52\pm0.05$) and the fit of 
Kaspi et al. (2005; {\it dashed}) 
agree in this case.  The intrinsic scatter (Table 2) is $\sim 0.3$ dex. 
NGC 5548 is included twice, and indicated with a red box in all panels.
({\bf b}) As in ({\bf a}), but here using the \hbeta\ rather than
X-ray luminosity.  In this case, our maximum-likelihood fit ({\it
  solid}; $\beta=0.53\pm0.04$) is significantly shallower than that of
Kaspi et al., bringing the slope into agreement with that of
\citet{bentzetal2009rl} for the optical continuum.  For reference, we show 
all fifteen epochs of monitoring for NGC 5548 in gray.
({\bf c}) The \rblr--\loiii\ relation.  As above, LAMP sources are 
open circles while the non-LAMP sources are small black circles. The solid
line is our best maximum-likelihood fit ($\beta=0.60\pm0.07$).
({\bf d}) As in ({\bf c}), but here using the [O {\tiny IV}]$~25.8~\micron$ 
luminosity rather than the [O {\tiny III}] luminosity.
The arrow indicates the upper limit on the [O {\tiny IV}] luminosity
of PG$2130+099$. Our maximum-likelihood fit ($\beta=0.69\pm0.13$) 
is shown as a solid line.
\label{fits}
}
\end{figure*}

We perform a second check using artificial light curves.  Our
goal is to investigate whether systematic changes in break time-scale
will lead to a bias in our derived \rblr--$L_{\rm 2-10 keV}$ relation.
We use the prescriptions of \citet{timmerkoenig1995} to generate mock
light curves with an input power spectrum of variability.  We generate
a family of light curves, each of 5 yr duration, and each with a
characteristic break in the power spectral density function. The break
timescales range from 0.01 to 30 days, which is similar to the range
of 0.005 to 30 days seen in reverberation-mapped sources
\citep[e.g.,][]{uttleymchardy2005}.  For simplicity, all light curves
have a power-law slope of $\alpha=-2$ ($P \propto f^{\alpha}$) at
frequencies above the break frequency and a slope of $\beta=-1$ ($P
\propto f^{\beta}$) at frequencies below the break.  Each artificial
light curve is ``observed'' 2000 times with 10~ks duration.  The
signal-to-noise ratio is taken to be 100 (but does not impact the results) 
and the assumed variability
amplitude is $20\%$ to match the Tartarus average above.  We then
look at the spread in derived X-ray fluxes as a function of break
timescale.  There is no change in the width of the distribution of
mean fluxes for breaks ranging from 0.01 to 30 days.  Therefore, we do
not expect any systematic errors in our X-ray flux estimates as a
function of mass or luminosity based on trends between break timescale
and mass or Eddington ratio \citep{mchardyetal2006}.

It is also worth discussing the primary sources of systematic errors
in the fluxes of narrow emission lines.  First, aperture corrections
can be significant, since narrow-line regions (NLRs) have sizes of
hundreds of parsecs and are often spatially resolved
\citep[e.g.,][]{whittle1992a,bennertetal2002,schmittetal2003obs,
  schmittetal2003res,greeneetal2009}.  Objects that are closer are
more susceptible to aperture effects.  The LAMP targets were observed
with a slit of width $4\arcsec$, while the non-LAMP targets generally
come from apertures of width 2--20\arcsec\ (although most are larger
than 4\arcsec), and the \citet{borosongreen1992} observations were
taken with a $1\farcs5$ slit.  Between the compilations of
\citet{bennertetal2002} and \citet{schmittetal2003obs}, we find NLR
sizes for four of the PG quasars, two LAMP objects (NGC 5548 and Mrk
766), and five other non-LAMP objects (Mrk 590, NGC 3516, Mrk 79, NGC
3783, NGC 4593).  The NLR sizes of these galaxies are 0.1--2\arcsec,
with a median size of 0\farcs3.  The PG quasars all have NLR sizes
$<$1\farcs2.  Nominally, based on the objects with measured NLR sizes,
we expect minimal loss of light due to aperture effects.  Observing
conditions will also lead to some slit losses, in the case of the
\citet{borosongreen1992} observations, but not at a level that is
significant compared to internal extinction (see below). To address
the rest of the sample, we estimate NLR sizes using the
size-luminosity relation of \citet{schmittetal2003res}.  We find that
the typical expected NLR size (for both LAMP and non-LAMP sources) is
$\sim 1\farcs3$.  The PG quasars, which were observed with the
smallest slit, all have expected sizes in the range
$0\farcs5-1$\arcsec.  We are in even less danger with the \loiv\
measurements, given their larger aperture sizes (5--20\arcsec) and
compact emission regions \citep{melendezetal2008}.

The next important concern is internal extinction.  It has been 
shown many times, particularly for obscured AGNs, that the \oiii\ 
luminosity can be significantly extinguished by dust
\citep[e.g.,][]{malkan1983,mulchaeyetal1994,netzeretal2006,melendezetal2008}.
On the other hand, \loiv\ is relatively insensitive to extinction,
making it a higher fidelity luminosity indicator
\citep[e.g.,][]{melendezetal2008,diamondstanicetal2009}; if the
\loiii\ measurements are compromised, we still expect to find
reasonable results for \loiv.  Finally, there is the possibility of
contamination from star formation.  In principle this is possible for
\loiii, but note that the \oiii/\hbeta\ intensity ratio is considerably 
lower in high-metallicity star-forming galaxies than in active 
galaxies \citep[e.g.,][]{baldwinetal1981}.  As shown by
\citet{kauffmannetal2003agn}, the expected level of \loiii\
contamination from star formation in local obscured active galaxies is
low ($<10\%$).

\section{Fits}

Our primary goal is to calibrate the relation between BLR size and
various indicators of nonstellar (AGN)
 luminosity.  We fit to the standard relation
($R_{\rm BLR}/10\,$lt-days) $= \alpha + \beta\, {\rm log}\, L$, where $L$ here is
derived from \hbeta, $L_{\rm 2-10 keV}$, and narrow emission-line
luminosities.  For comparison with recent literature, we
utilize two primary fitting schemes.  The first is a \chisq\
minimization technique similar to that presented by
\citet{tremaineetal2002}.  The following \chisq\ function is
minimized:
\begin{equation}
\chi^2 \equiv \sum_{i=1}^{N} \frac{(R_i - \alpha - \beta L_i)^2}
{\epsilon^2_{Ri} + \beta^2 \epsilon^2_{Li}}.
\end{equation}
Intrinsic scatter is accounted for by replacing 
$\epsilon_{Ri}$ with $\epsilon=(\epsilon^2_{Ri} + \epsilon^2_0)^{1/2}$,
where $\epsilon_0$ (the intrinsic scatter) is chosen such that $\chisqr 
= 1$.

In addition, we use a maximum-likelihood technique adapted from
\citet{gultekinetal2009}.  For simplicity we assume that both the
measurement errors and the intrinsic scatter have Gaussian
distributions.  For a set of observed points ($R_{i}, L_{i}$), we
maximize the total likelihood,
\begin{equation}
\mathcal{L} = \prod_i l_i (R_{i}, L_{i}).  
\end{equation}
In the presence of measurement errors, if the likelihood of measuring
a BLR radius $R_{i}$ for a true radius $R$ is $Q_i(R_{i}\vert R)\,
dR_{i}$, and the probability to have a true radius $R$ given $L_{i}$
is $P$, then for a given observation the likelihood is
\begin{equation}
l_i = \int Q_i (R_i \vert R) P (R \vert L_{i})\, dL.
\end{equation}
We assume that both $Q$ and $P$ have a log-normal form.  Uncertainties 
in the independent variable (luminosity) are derived from Monte Carlo 
simulations and are always small. Fits using both
methods are given in Table 2, with the first line showing the \chisq\
method. In all cases the results of the two fitting methods are indistinguishable.

It is interesting to note that the X-ray and \hbeta\ relations are now
consistent with a slope of $R_{\rm BLR} \propto \sqrt{L}$.  In
contrast, \citet{kaspietal2005} report a slope of 0.7 for the X-ray
relation ($R_{\rm BLR} \propto L_{\rm 2-10 keV}^{0.7}$).  We should note,
however, that when they fit an average lag for each object and used
only \hbeta\ lags (the most directly comparable case to what we have
done here), they find a slope of 0.5 ($R_{\rm BLR} \propto L_{\rm 2-10
  keV}^{0.53}$).  Their reported $R_{\rm BLR}-L_{{\rm H}\beta}$ slope is
steeper, $R_{\rm BLR} \propto L_{\hbeta}^{0.69}$.  With our improved
data, we find that both relations are consistent with a slope of 0.5.
Thus, the simplest assumption, that AGN SEDs and BLR densities are
independent of luminosity, appears to apply, at least for the present
sample and to the level of precision that can be tested by our data.
One goal of ongoing reverberation-mapping campaigns should be to
investigate whether there are physical regimes (e.g., in BH mass or
luminosity) for which this assumption does not hold
\citep[e.g.,][]{greeneho2009}.

There is tantalizing evidence, in contrast, that the narrow-line
relations may have a steeper slope, although with low significance.
Here we explore possible interpretations of this result, should it
turn out to be significant.  Above we discussed various sources of
contamination of the NLR luminosity, including redshift-dependent
aperture correction, extinction, and star formation.  Aperture effects
go in the wrong direction to explain the steeper slope, while
extinction seems implausible because it would have to impact the \loiv\
measurements as strongly than the \loiii\ measurements, contrary
to normal reddening laws. Star
formation could artificially boost the NLR luminosities at the low
end.  However, we do not believe the \loiii\ contamination could be
more than $\sim 10\%$ on average, while the values need to be boosted
by factors of 2--3 to impact the slope on a logarithmic scale.
Therefore, the steeper slope, if real, is more likely explained by
physical effects rather than measurement errors.  It would most
naturally arise from the measured luminosity dependence in the
relation between NLR and bolometric luminosity.  We are not the first
to report this trend.  For instance, \citet{netzeretal2006} find
that $L_{\rm [O~III]} \propto L_{\rm 2-10~ keV}^{0.70\pm0.06}$, while
\citet{melendezetal2008} find $L_{\rm [O~IV]} \propto L_{\rm 
2-10~ keV}^{0.7\pm0.1}$ and $L_{\rm [O~III]} \propto L_{\rm 
[O~IV]}^{0.9\pm0.1}$. If \rblr $\propto$ \lhard$^{0.5}$, then based on
Netzer et al. we would expect \rblr $\propto$ \loiii$^{0.7 \pm 0.1}$,
which is consistent with our finding.  The slope we measure in the
\rblr--\loiv\ relation is also consistent with the results of
\citet{melendezetal2008}.  Thus, the possibility of a steeper
slope is plausible.  For some reason, quasars are less efficient 
at powering an NLR than are the less luminous Seyfert nuclei.

There is now compelling evidence that bolometric corrections depend on
the Eddington ratio \lledd, where the Eddington luminosity for
1~\msun\ is taken to be $1.25 \times 10^{38}$~erg~s$^{-1}$
\citep[e.g., ][]{vasudevanfabian2007,vasudevanfabian2009}. It is worth
seeking correlations between radius-luminosity relation residuals and
the Eddington ratio. BH mass measurements are provided by the
reverberation-mapping campaigns.  In the case of the non-LAMP sources,
we take the Eddington ratios from the study of
\citet{vasudevanfabian2009}, who have measured simultaneous SEDs
ranging from the optical to the X-ray using \xmm.  We do not yet have
full SEDs for the LAMP sample, and so we use a single-band 
observation and a bolometric correction.  We adopt \lhard\ and the
bolometric correction from \citet{vasudevanfabian2009}.  The
bolometric correction depends on the Eddington ratio and we assume a 
value of 30, as appropriate for sources with \lledd $\approx 10\%$.  
The resulting Eddington ratios are in the range 0.001--1, but are 
strongly peaked at $\sim 0.1$.

We seek correlations between the residuals in \rblr\ around the mean
\rblr--$L$ relations and the Eddington ratio. The nonparametric
Kendall's $\tau$ is calculated (within 
IRAF\footnote{http://iraf.noao.edu/.}) for all relations. In no case
do we find evidence for a correlation between the \rblr--$L$ residuals
and \lledd.  The probability of no correlation is in the range
$P=0.3-0.8$.

Although we do not know its origin, it is interesting to examine the
intrinsic scatter for each fit. As expected, the intrinsic scatter is
lowest when \lhb\ is used, presumably because of both temporal and
spatial coincidence.  On the other hand, the relations based on both
the X-rays and narrow emission lines have comparable scatter.  One
might expect higher scatter in the narrow emission-line relation due
to the unquantified role of internal extinction and aperture effects.
Furthermore, the narrow-line emission cannot respond to changes in
accretion luminosity on timescales of a month, while we know that
\rblr\ does.  Note that NGC 5548 has shown \rblr\ variability at the
factor of four level, and yet the overall relations only have an
intrinsic scatter of a factor of two.  Once the \lf\ measurements are
in hand, it will be interesting to see whether the intrinsic scatter
is minimized using the optical continuum luminosity or, indeed, the
bolometric luminosity.

\section{Summary}

We explore radius-luminosity relations based on AGN luminosities other
than the optical continuum.  The time is right to revisit these
relations because of a new sample of low-luminosity, low-mass AGNs
with reverberation mapping from the LAMP project
\citep{bentzetal2009lamp}.  We consider X-ray, broad \hbeta, narrow
\oiii, and narrow \oiv\ luminosities.  These relations are designed
for use in estimating BH masses when optical continuum luminosities
are not available.  Relevant situations include local AGNs with
galaxy-dominated spectra, and possibly radio-loud objects, various
high-redshift active galaxy populations (such as submillimeter
galaxies), and heavily obscured AGNs with detected broad polarized
emission.  Furthermore, any differences in slope or intrinsic scatter
between relations based on different luminosities may indicate SED
differences in AGNs as a function of luminosity or BH mass.

We find that the \rblr--\lhard\ and \rblr--\lhb\ relations are well fit
with a slope of \rblr $\propto \sqrt{L}$.  This is the slope expected
if AGN SEDs and BLR densities are independent of luminosity.  On the
other hand, the narrow emission lines show tentative evidence for
a steeper relation, \rblr $\propto L^{0.6}$.  Intriguingly, these
slopes are consistent with previous results showing that \loiii/\lx\
and \loiv/\lx\ decreases with increasing luminosity
\citep[e.g.,][]{netzeretal2006,melendezetal2008}.  We find no evidence
for a correlation between \rblr--$L$ residuals and Eddington ratio.  In
fact, the intrinsic scatter in all relations is surprisingly small.
On the one hand, the X-rays are variable on short timescales, but, as
we show, that does not translate into significant errors in the
\rblr--\lhard\ relation.  On the other hand, the narrow emission-line
luminosities do not respond at all to state changes on timescales of
a year.  Thus, we find it surprising that even in these cases the
intrinsic scatter is only at the factor of two level.  Still, this
scatter translates directly into uncertainties in the BH masses
\citep[e.g.,][]{vestergaardpeterson2006,mcgilletal2008}. As the 
reverberation-mapped samples increase, it should  become possible
to search for evidence of secondary parameters that might allow one
to decrease the total scatter, thereby increasing the fidelity of our BH
mass estimates.

\acknowledgements{
The referee gave many valuable comments that substantially
improved this manuscript.  We thank the excellent staff and support
personnel at Lick Observatory for their enormous help during our
observing run, and L.~C.~Ho for inspiring conversations.  This work
was supported by NSF grants AST-0548198 (UC Irvine), AST-0607485 and
AST-0908886 (UC Berkeley), AST-0642621 (UC Santa Barbara), and
AST-0507450 (UC Riverside). The UC Berkeley researchers also
gratefully acknowledge the support of both the Sylvia \& Jim Katzman
Foundation and the TABASGO Foundation for the continued operation of
the Katzman Automatic Imaging Telescope (KAIT), with which some of the
photometry was obtained. M.C.B. gratefully acknowledges support 
provided by NASA through Hubble Fellowship grant HF--51251 awarded by 
the Space Telescope Science Institute, which is operated by the 
Association of Universities for Research in Astronomy, Inc., for NASA, 
under contract NAS 5-26555.
This research has made use of the Tartarus (Version 3.1) database,
created by Paul O'Neill and Kirpal Nandra at Imperial College
London, and Jane Turner at NASA/GSFC. Tartarus is supported by
funding from PPARC, as well as from NASA grants NAG5-7385 and NAG5-7067.}

\end{document}